# Algorithm and approaches to handle large Data- A Survey

[1]Chanchal Yadav, [2]Shuliang Wang , [3]Manoj Kumar

[1] CSE, Amity University
Noida, Uttar Pradesh, India
*Chanchalyadav90@gmail.com*

[2]School of Software,
Beijing Institute of Technology
Beijing, China
*slwang2011@bit.edu.cn*

[3] CSE, Amity University
Noida, Uttar Pradesh, India
*manojbaliyan@gmail.com*

**Abstract**

Data mining environment produces a large amount of data, that need to be analyzed, patterns have to be extracted from that to gain knowledge. In this new era with boom of data both structured and unstructured, in the field of genomics, meteorology, biology, environmental research and many others, it has become difficult to process, manage and analyze patterns using traditional databases and architectures. So, a proper architecture should be understood to gain knowledge about the Big Data. This paper presents a review of various algorithms from 1994-2013 necessary for handling such large data set. These algorithms define various structures and methods implemented to handle Big Data, also in the paper are listed various tool that were developed for analyzing them.

## 1. Introduction

Data Mining is the technology to extract the knowledge from the data. It is used to explore and analyze the same. The data to be mined varies from a small data set to a large data set i.e. big data.
Data Mining has also been termed as data dredging, data archeology, information discovery or information harvesting depending upon the area where it is being used. The data Mining environment produces a large volume of the data. The information retrieved in the data Mining step is transformed into the structure that is easily understood by its user. Data Mining involves various methods such as genetic algorithm, support vector machines, decision tree, neural network and cluster analysis, to disclose the hidden patterns inside the large data set.

Since Data Mining environment produces a large amount of data, to classify this large amount of data, to extract patterns and classify data with high similar traits, Data

Mining approaches such as Genetic algorithm, neural networks, support vector Machines, association algorithm, clustering algorithm, cluster analysis, were used. In other words, we may also Say that, Data Mining is the process of recovering patterns among many fields in the database.
Big Data are the large amount of data being processed by the Data Mining environment. In other words, it is the collection of data sets large and complex that it becomes difficult to process using on hand database management tools or traditional data processing applications, so data mining tools were used. Big Data are about turning unstructured, invaluable, imperfect, complex data into usable information. [1]
Data have hidden information in them and to extract this new information; interrelationship among the data has to be achieved. Information may be retrieved from a hidden or a complex data set.
Browsing through a large data set would be difficult and time consuming, we have to follow certain protocols , a proper algorithm and method is needed to classify the data, find a suitable pattern among them. The standard data analysis method such as exploratory, clustering, factorial, analysis need to be extended to get the information and extract new knowledge treasure.
Due to Increase in the amount of data in the field of genomics, meteorology, biology, environmental research and many others, it has become difficult to find, analyze patterns, associations within such large data .
Interesting association is to be found in the large data sets to draw the pattern and for knowledge purpose.
The benefit of analyzing the pattern and association in the data is to set the trend in the market, to understand customers, analyze demands, predict future possibilities in every aspect. It helped organization to increase innovation, retain customers, and increase in operational efficiency. [2] Big Data can be measured using the following that is categorized by 4 V's: variety, volume, velocity and value. [2][3]

The following paper reviews different aspects of the big data. The paper has been described as follows, Section I: Introduction about Big data and application of big data. Section II: deals with the architecture of the big data. Section III: describes the various algorithms used to process Big Data. Section IV: describes potential applications of Big data. Section V: deals with the various algorithms specifying ways to handle big data. Section VI: deals with the various security issues related to the big data.

## 2. Architecture

Big Data are the collection of large amounts of unstructured data. Big Data means enormous amounts of data, such large that it is difficult to collect, store, manage, analyze, predict, visualize, and model the data.

Big Data architecture typically consists of three segments: storage system, handling and analysis.

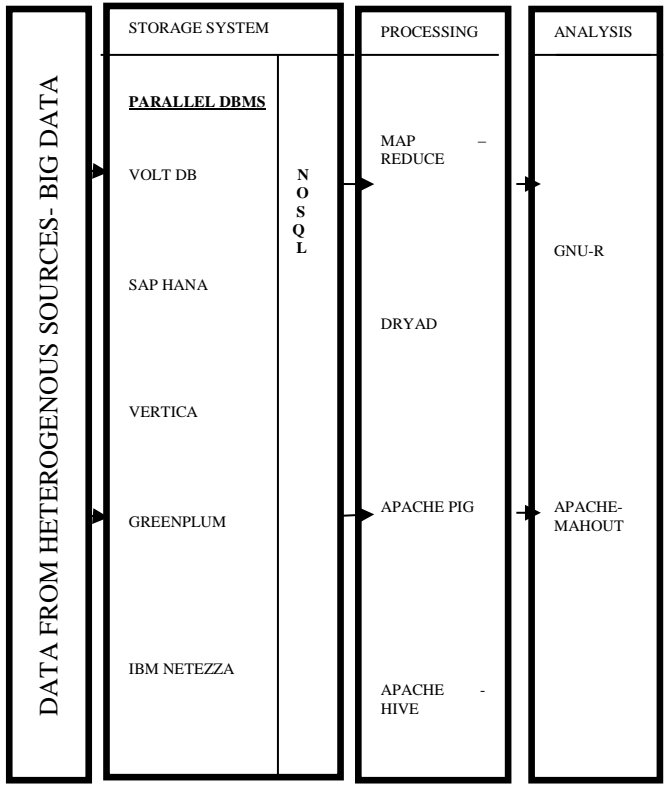

Fig. 1 Architecture of Big Data

Big Data typically differ from data warehouse in architecture; it follows a distributed approach whereas a data warehouse follows a centralized one.

One of the architecture laid describes about adding new 6 rules were in the original 12 rules defined in the OLAP system defined the methods of data mining required for the analysis of data and defined SDA (standard data analysis) that helped analysis of data that is in aggregated form and these were much well timed in comparison with the decision taken in traditional methods. [4]

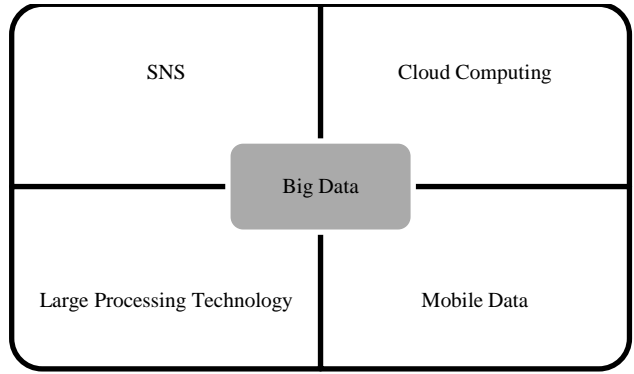

Fig. 2 Big Data modes

On paper An Efficient Technique on Cluster Based Master Slave Architecture Design, the hybrid approach was formed which consists of both top down and bottom up approach. This hybrid approach when compared with the clustering and Apriori algorithm, takes less time in transaction than them. [5]

The Data Mining termed Knowledge discovery, in work done in Design Principles for Effective Knowledge Discovery from Big Data", its architecture was laid describing extracting knowledge from large data. Data was analyzed using software Hive and Hadoop. For the analysis of data with different format cloud structure was laid. [6]

Then there arrived the prime need to analyze the unstructured data, so the Hadoop framework was being laid for the analysis of the unstructured large data sets. [7]

## 3. Algorithm

Many algorithms were defined earlier in the analysis of large data set. Will go through the different work done to handle Big Data. In the beginning different Decision Tree Learning was used earlier to analyze the big data. In work done by Hall. et al. [10], there is defined an approach for forming learning the rules of the large set of training data. The approach is to have a single decision system generated from a large and independent n subset of data. Whereas Patil et al, uses a hybrid approach combining both genetic algorithm and decision tree to create an optimized decision tree thus improving efficiency and performance of computation. [12].

Then clustering techniques came into existence. Different clustering techniques were being used to analyze the data sets. A new algorithm called GLC++ was developed for large mixed data set unlike algorithm which deals with large similar type of dataset. This method could be used with any kind of distance, or symmetric similarity function. [13]

| AUTHOR'S NAME | TECHNIQUE | CHARACTERISTIC | SEARCH TIME |
|---|---|---|---|
| N. Beckmann, H. -P. Kriegal, R. Schneider, B. Seeger [8] | R-Tree R*-Tree | Have performance bottleneck | $O(3^D)$ |
| S. Arya, D. Mount, N. Netanyahu, R. Silverman, A. Wu [9] | Nearest Neighbor Search | Expensive when searching object is in High Dimensional space | Grows exponentially with the size of the searching space. $O(dn \log n)$ |
| Lawrence O. Hall, Nitesh Chawla, Kevin W. Bowyer [10] | Decision Tree Learning | Reasonably fast and accurate | Less time consuming |
| Zhiwei Fu, Fannie Mae [11] | Decision Tree C4.5 | Practice local greedy search throughout dataset | Less time consuming |
| D. V. Patil, R. S. Bichkar [12] | GA Tree (Decision Tree + Genetic Algorithm) | Improvement in classification, performance and reduction in size of tree, with no loss in classification accuracy | Improved performance-Problems like slow memory, execution can be reduced |
| Yen-Ling Lu, Chin-Shyurng Fahn [17] | Hierarchical Neural Network | High accuracy rate of recognizing data; have high classification accuracy | Less time consuming-improved performance |

Table 1: Different Decision tree Algorithm

Whereas Koyuturk et al. Defined a new technique PROXIMUS for compression of transaction sets, accelerates the association mining rule, and an efficient technique for clustering and the discovery of patterns in a large data set. [14].

With the growing knowledge in the field of big data, the various techniques for data analysis- structural coding, frequencies, co-occurrence and graph theory, data reduction techniques, hierarchical clustering techniques, multidimensional scaling were defined in Data Reduction Techniques for Large Qualitative Data Sets. It described that the need for the particular approach arise with the type of dataset and the way the pattern are to be analyzed. [15] The earlier techniques were inconvenient in real time handling of large amount of data so in Streaming Hierarchical Clustering for Concept Mining, defined a novel algorithm for extracting semantic content from large dataset. The algorithm was designed to be implemented in hardware, to handle data at high ingestion rates. [16]. Then in Hierarchical Artificial Neural Networks for Recognizing High Similar Large Data Sets., described the techniques of SOM (self-organizing feature map) network and learning vector quantization (LVQ) networks. SOM takes input in an unsupervised manner whereas LVQ was used supervised learning. It categorizes large data set into smaller thus improving the overall computation time needed to process the large data set. [17]. Then improvement in the approach for mining online data come from archana et al. Where online mining association rules were defined to mine the data, to remove the redundant rules. The outcome was shown through graph that the number of nodes in this graph is less as compared with the lattice. [18]. Then after the techniques of the decision tree and clustering, there came a technique in Reshef et al. In which dependence was found between the pair of variables. And on the basis of dependence association was found. The term maximal information coefficient (MIC) was defined, which is maximal dependence between the pair of variables. It was also suitable for uncovering various non-linear relationships. It was compared with other approaches was found more efficient in detecting the dependence and association. It had a drawback –it has low power and thus because of it does not satisfy the property of equitability for very large data set. [19]. Then in 2012 wang, uses the concept of Physical Science, the Data field to generate interaction between among objects and then grouping them into clusters. This algorithm was compared with K-Means, CURE, BIRCH, and CHAMELEON and was found to be much more efficient than them. [20]. Then, a way was described in "Analyzing large biological datasets with association network" to transform numerical and nominal data collected in tables, survey forms, questionnaires or type-value annotation records into networks of associations (ANets) and then generating Association rules (A Rules). Then any visualization or clustering algorithm can be applied to them. It suffered from the drawback that the format of the dataset should be syntactically and semantically correct to get the result. [21]. Later, in A Survey of Different Issues of Different Clustering Algorithms used in Large Data Sets classifies different clustering algorithms and gives an overview of different clustering algorithm used in large data sets. [22].

## 4. Potential Application

Big Data is very beneficial when it comes to organization. It helps them to compete more effectively with other organization, better understanding of the customer, grow business revenue, and others. It has been found giving effective results in the field of: retail, banking, insurance, government, natural resource, healthcare, manufacturing, public sector administration, personal data location and other services. [25][26][27][29]

## 5. Issues with Big Data

Security has always been an issue when data privacy is considered. Data integrity is one of the primary components when preservation of data is considered. Access and sharing of Data which is not meant for public, has to be protected. For this type of security many researchers have been done. Security has always been an issue when data are considered. In the paper A Metadata Based Storage Model for Securing Data In Cloud Environment, defined the metadata based approach to secure the large data. They provide the architecture to store the data. Uses cloud computing to make the data unavailable to the intruder. [23] Data integrity is one of the primary components when preservation/security is considered. Hash functions were primarily used for preserving the integrity of the data. The drawback of using hash function is that a single hash can only identify the integrity of the single data string. And because of this drawback, it becomes impossible to locate the exact position within the string where the change has been occurring. The solution to overcome the above problem is to split the data string into the block and then protect each block by the hash function. This also created a drawback that in case of large data set storing such large number of hashes imposes significant space overhead. In paper Hashing Scheme for Space-efficient Detection and Localization of Changes in Large Data Sets, method to overcome this problem was described. Certain properties like logarithmic were added instead of linear increase. [24]. Whereas the work explained by paper Big Data Privacy Issues in Public Social Media, the very idea of privacy to the people who are using social media was explained. The 3 techniques to get the location information to stay away from such harmful flood of information were explained. [25]

## 6. Conclusion

Due to Increase in the amount of data in the field of genomics, meteorology, biology, environmental research, it becomes difficult to handle the data, to find Associations, patterns and to analyze the large data sets. As organizations continue to collect more data at this scale, formalizing the process of big data analysis will become paramount.
The paper describes different methodologies associated with different algorithms used to handle such large data sets.
And it gives an overview of architecture and algorithms used in large data sets. It also describes about the various security issues, application and trends followed by a large data set.

**Chanchal Yadav,** *M-tech (pursuing), Amity School of engineering and technology, Amity University, Noida, U.P., India*

**Shuliang Wang,** *PhD, is a professor at the Beijing Institute of Technology in China. His research interests include spatial data mining and software engineering. For his innovatory study of spatial data mining, he was awarded one of the best national thesis, the Fifth Annual InfoSciR-Journals Excellence in Research Awards, IGI Global and so on.*